\documentclass[
               ,draft 
              ]{aipproc}
\layoutstyle{6x9}

\usepackage{natbib}

\begin{document}


\title{The Ejection of Low Mass Clumps During Star Formation}

\author{Shantanu Basu}
       {address={Department of Physics \& Astronomy, Western University, London, Ontario, N6A 3K7, Canada}}
\author{Eduard I.~Vorobyov}
       {address={Institute for Astrophysics, The University of Vienna, Vienna, 1180, Austria},altaddress={Institute of Physics, Southern Federal University, Stachiki 194, Rostov-on-Don, 344090, Russia}}
\author{Alexander L.~DeSouza}
       {address={Department of Physics \& Astronomy, Western University, London, Ontario, N6A 3K7, Canada}}

\begin{abstract}
Modeling of the self-consistent formation and evolution of disks as a result 
of prestellar core collapse reveals an intense early phase of recurrent 
gravitational instability and clump formation. These clumps generally 
migrate inward due to gravitational interaction with trailing spiral arms, and 
can be absorbed into the central object. However, in situations of multiple 
clump formation, gravitational scattering of clumps can result in the ejection
of a low mass clump. These clumps can then give rise to free-floating low mass
stars, brown dwarfs, or even giant planets. Detailed modeling of this process 
in the context of present-day star formation reveals that these clumps 
start out essentially as Larson first cores and grow subsequently by accretion.
In the context of Pop III star formation, preliminary indications are that the
disk clumps may also be of low mass. This mechanism of
clump formation and possible ejection provides a 
channel for the formation of low mass objects in the first generation of 
stars.
  
\end{abstract}



\maketitle


\section{Introduction}
Numerical simulations that can track the evolution all the way
from prestellar core collapse through the self-consistent formation of a disk
and its subsequent long-term evolution are revealing new insights into 
star, brown dwarf, and planet formation. Our group has published a series of
papers \citep[e.g.,][]{vor06,vor10,bas12} that illustrate an early phase of
disk evolution that is characterized by recurrent gravitational instability, and
accretion driven by gravitational torques. Gravitational instability is triggered
when accretion onto the disk drives the Toomre parameter of the disk below the
critical value, resulting in the formation of gas clumps within nonlinear spiral 
arms. These clumps are generally driven inward through gravitational torques
resulting from their interaction with trailing spiral arms. Some clumps are also
dispersed due to tidal effects. Clumps that plunge inward to the central object
can be invoked to explain the (FU Ori) luminosity bursts that are associated with 
young stellar objects \citep{ken90}. Recent modeling \citep{bas12} shows that the collapse
of cores with sufficient mass and/or angular momentum can lead to disks with 
multiple clump formation in which the gravitational scattering of clumps leads
to the ejection of low mass clumps. These clumps generally straddle the substellar
limit, and may be the precursors of free-floating low mass stars, brown dwarfs, or
even giant planets. 

The above modeling is done in the thin-disk approximation, with a central sink
cell of about 6 AU in radius, and a logarithmically spaced radial grid. 
These simplifications allow the modeling of the long-term evolution ( > 1 Myr) of 
the disk and core envelope structure.
The combination of these features is generally not both feasible in fully three-dimensional simulations,
even with adaptive mesh refinement. However, new three-dimensional simulations that
resolve a large dynamic range of length scales have recently also found some of the same
features, e.g., gravitational instability and episodic accretion \citep{mac11}.


\section{Model \& Results}

Figure 1 shows a sequence of column density images for a reference model 
presented in \citet{bas12}, at various times after the formation of a central
object, on the full computational domain of 20,000 AU on each side.
Fragments are formed as early as 0.05 Myr, but here we focus in on a 
narrow time window within which an ejection event occurs. 
Although clumps are generally torqued inward or sometimes disperse, 
under favorable conditions a clump within a multi-clump environment may
be ejected through many-body interaction. The ejection is also aided by the
nonaxisymmetric potential of the disk. The arrows point to a clump that
undergoes an ejection. The speed of the ejected clump at the moment that it
leaves the computational domain is about 0.9 km s$^{-1}$, which is a factor of
three greater than the escape speed from the system. However, it is comparable
to typical random speeds of young stellar objects within their host clusters.
The total mass of the clump, calculated as the mass passing the computational 
boundary during the ejection event, is 0.15 $M_{\odot}$. We
speculate that upon contraction this clump may form a substellar
object, given that a significant fraction of mass remains in the
disk until it is ejected due to outflows and/or dispersed due to
photoevaporation. Other models presented by \citet{bas12} show the ejection 
of even substellar mass objects. 


\section{Discussion and Conclusions}

The model presented here has important implications for the formation of 
free-floating low mass stellar or substellar objects in the Galaxy. But what
are the implications for the first generation of stars?

The clumps in the models with solar metallicity start out as essentially
Larson first cores, defined as objects of a Jeans mass at the density
where optical depth unity is first reached. This means that their initial mass 
is about $0.01\, M_{\odot}$ and they can grow subsequently by accretion from 
the disk, usually achieving
a mass of about ten times this value, or $\sim 0.1\, M_{\odot}$.
Figure 2 shows the temperature-density relations for both $Z=Z_{\odot}$ and 
$Z=0$ based on calculations of \citet{omu05}. While optical depth unity is reached
at a number density of $n \simeq 10^{11}$ cm$^{-3}$ for solar metallicity, it is
reached at $n \simeq 10^{16}$ cm$^{-3}$ for the $Z=0$ case \citep{omu05}. This
is also the location of a local steepening of the temperature-density relation.
The Jeans mass at this density and corresponding temperature $\simeq 2000$~K is 
$\sim\, {\rm few}\, \times 0.01\, M_{\odot}$. 
Preliminary calculations \citep{vor12,des12} do find the formation of disk
fragments of mass $\sim 1\, M_{\odot}$, which is roughly consistent with the 
present-day simulations where the clumps grow to about ten times the initial 
Jeans mass. The formation, growth, and possible ejection of low mass Pop III stars 
through disk fragmentation is thus a tantalizing possibility that can be clarified
by future calculations.

\begin{figure}[h!]
  \includegraphics[width=0.8\textwidth]{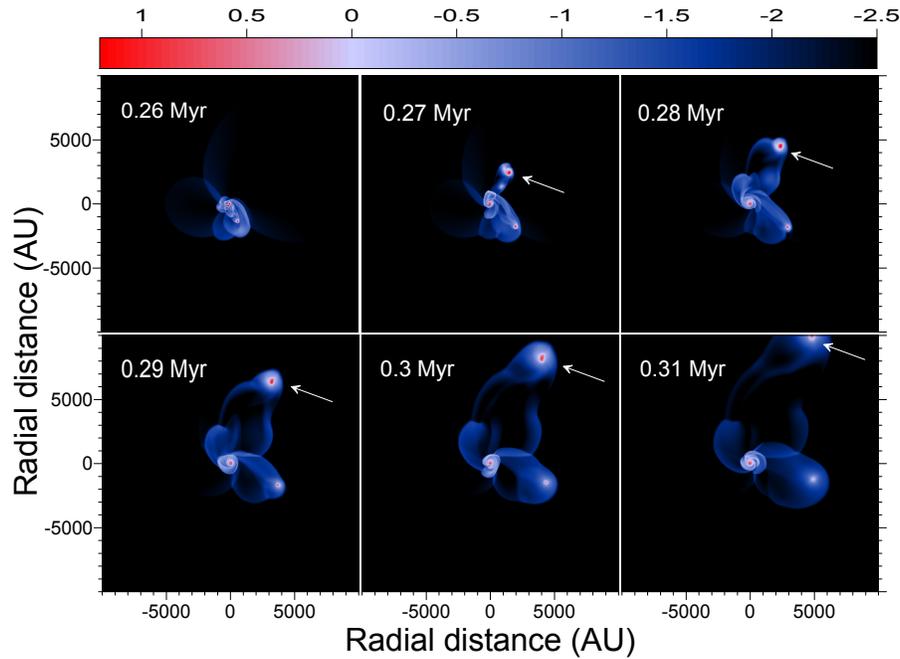}
\caption{Gas surface density distribution (g cm$^{-2}$, log units) in a reference model at several time instances after the formation of a central star. The box size is
20,000 AU on each side and represents nearly the full extent of the computational domain. Arrows identify a clump that is ejected from the system after a multi-body interaction within the centrifugal disk.}
  \label{fig:ejection}
\end{figure}

\begin{figure}[h!]
  \includegraphics[width=0.8\textwidth]{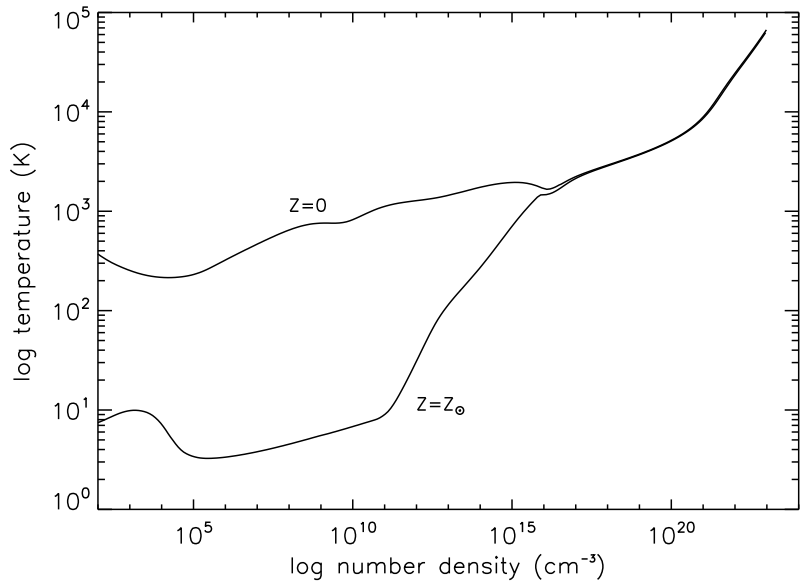}
  \caption{Temperature-density evolution for present day ($Z=Z_{\odot}$) and primordial 
  ($Z=0$) metallicities, based on the detailed thermal balance calculations of \citet{omu05}}
  \label{fig:barotropic}
\end{figure}


\begin{theacknowledgments}
We thank the organizers of First Stars IV for their efforts in putting together a highly stimulating conference. EIV acknowledges support from the RFBR grants 10-02-00278 and
11-02-92601. SB was supported by an NSERC Discovery Grant.
\end{theacknowledgments}


\end{document}